\documentclass[epj]{svjour}

\usepackage{amssymb}
\usepackage{graphicx}

\newcommand{\ellt}{\tilde{\ell}}

\begin{document}

\title{Variational theory for a single polyelectrolyte chain revisited}

\author{Manoel Manghi\thanks{\email{manghi@theorie.physik.uni-muenchen.de}} \and
Roland R. Netz\thanks{\email{netz@theorie.physik.uni-muenchen.de}}}

\institute{Sektion Physik, Ludwig Maximilians University, Theresienstr. 37, 80333
Munich, Germany}

\date{Received: date / Revised version: date}


\abstract{We reconsider the electrostatic contribution to the persistence
length, $\ell_e$, of a single, infinitely long charged polymer in the presence
of screening. A Gaussian variational method is employed, taking $\ell_e$ as the
only variational parameter. For weakly charged and flexible chains, crumpling
occurs at small length scales because conformational fluctuations overcome
electrostatic repulsion. The electrostatic persistence length depends on the
square of the screening length, $\ell_e\sim\kappa^{-2}$, as first argued by
Khokhlov and Khachaturian by applying the Odijk-Skolnick-Fixman (OSF) theory to
a string of crumpled blobs. We compare our approach to previous theoretical
works (including variational formulations) and show that the result
$\ell_e\sim\kappa^{-1}$ found by several authors comes from the improper use of
a cutoff at small length scales. For highly charged and stiff chains, crumpling
does not occur~; here we recover the OSF result and validate the perturbative
calculation for slightly bent rods.}



\maketitle


\section{Introduction}


Charged polymer chains, also called polyelectrolytes, have many industrial
applications (flocculants, viscosifiers, adsorbants, etc.) related to their
large solubility in water, which is present even for a highly hydrophobic
backbone. They are also important in biology since many biopolymers (such as
nucleic acids) are charged.

In this paper we reconsider the controversial issue of the dependence of the
persistence length of poly\-electrolytes on electrostatic interactions and,
consequently, on the salt concentration in water, which has been of central
interest during the past few years~\cite{Forster,BarratRev,NetzRev}. This is
an important question with a large number of experimental implications, since
for instance, many  properties of biological and synthetic polymers depend on
their stiffness and thus on the salt concentration in the solution. It is
therefore an important control parameter. Most of the theoretical attempts have
treated electrostatic interactions on the linear Debye-H\"uckel level, leaving
aside the very important effects connected with counterion condensation (which
can be obtained on the mean-field level by properly taking non-linear effects
into account) and with strong electrostatic correlations, important when
multivalent ions are present~\cite{kardar,shklovskii,ariel}. Here we use the
same simplifying assumptions. Our purpose is therefore not to describe
experiments or simulations (with explicit counterions) better than previous
calculations, but to detect and eliminate inherent methodological shortcomings
and inconsistencies in some of the previous calculations, which led to
controversial results. Our results should therefore not be compared with
experiments, but rather with one of the many simulations of polymer chains where
Debye-H\"uckel repulsions are used~\cite{Micka,Stevens,Everaers,Nguyen,Ullner}.
Especially the disagreement between a number of previous variational
calculations is highly
irritating~\cite{Schmidt,Barrat,Bratko,Li,Thirumalai1,Netz,Thirumalai2}:  
variational calculations are typically known to be very robust, reliable and therefore
useful in a wide range of physics applications. The immediate question therefore
should be~: is there any ingredient in fluctuating charged polymers that forbids
the use of a variational theory~? The answer we give in this paper is clearly
negative. Variational methods do work for charged polymers as they do for
superconductivity, strongly correlated electron systems, protein folding, random
media, etc. This makes our methods in principle also applicable to more
complicated problems such as polymer solutions and the coupling between
counterion condensation and chain stiffening (which we will tackle in the
future).

Our calculational strategy consists in constructing the simplest variational
Gaussian correlation function which i)~reflects the stiffening of the chain at
small length scales, but at the same time, ii)~leads to a converging variational
free energy and therefore makes introduction of a small-scale cutoff (as
used in previous calculations) unnecessary. Quite surprisingly, the simple
worm-like chain correlation function does not satisfy criterion ii) and
therefore has to be discarded. The simplest workable correlation function,
related to the tangent-tangent correlation function, shows a two-scale
behaviour, in agreement with previous perturbative
calculations~\cite{BarratRev}. It explains why the electrostatic persistence
length is difficult to extract in experiments or simulations. As our main
result, we obtain that, in the flexible polymer case, where the chain crumples
at small length scales and forms a persistent chain of Gaussian blobs
(\textit{Gaussian-persistent} regime), the electrostatic persistence length
depends on the square of the screening length, $\ell_e\sim\kappa^{-2}$. We thus
confirm the scaling theory by Khokhlov and Khachaturian~\cite{KK} who applied
the Odijk-Skolnick-Fixman (OSF)~\cite{Odijk,SF} theory to a string of crumpled
blobs. This is in accordance with previous theoretical
studies~\cite{Netz,Thirumalai2} and very recent
simulations~\cite{Everaers,Nguyen}. The uncontrolled usage of a cutoff at small
length scales seems to be the origin of the very different results for the
persistence length found in previous papers. In the next section, we summarize
previous calculations for the persistence length of charged polymers. In
Section~3, we present the general scheme of our calculations both for flexible
and stiff chains. In Section~4, we study the case of flexible chains, using a
variational Hamiltonian which exhibits a crossover from crumpled statistics at
small scales to rod-like behaviour at intermediate scales. We show both
numerically and analytically in the asymptotic limit of low ionic strength that
$\ell_e\sim\kappa^{-2}$. In Section~5, we return to the case of stiff polymers,
the \textit{persistent} regime, and show why the worm-like-chain model cannot be
used per-se for a variational approach. This is related to the behaviour at
small scales. We use a slightly more complicated variational kernel which
regularizes the small-scale behaviour and recover variationally the classical
Odijk-Skolnick-Fixman result. Section~6 is devoted to the discussion of our
results and the comparison with previous theoretical calculations. Finally in
Section~7, we give our concluding remarks.


\section{Previous Results}


The first studies of the influence of electrostatic interactions on
polyelectrolyte stiffness have been done by Odijk~\cite{Odijk} and Skolnick
and Fixman~\cite{SF}. By performing a perturbation calculation on a slightly
bent rigid charged rod using the Debye-H\"uckel approximation, they found
that the total persistence length is
\begin{equation}
\ell_p=\ell_0+\ell_e
\end{equation}
where $\ell_0$ is the bare persistence length and $\ell_e$ the electrostatic
contribution:
\begin{equation}
\ell_e=\ell_{\mathrm{OSF}} =\frac{\ell_B}{4A^2\kappa^2}
\label{OSF}
\end{equation}
The length $\ell_B=e^2/(4\pi\epsilon k_BT)$ is the Bjerrum length (the distance
between two elementary charges interacting with the thermal energy $k_BT$), $A$
the distance between charges along the chain and $\kappa=[4\pi\ell_B
(c_++c_-)]^{1/2}$ the Debye-H\"uckel parameter, related to micro-ion
concentrations, $c_{\pm}$. This result is valid for polymer conformations which
do not deviate too much from the rod-like reference state, i.e. for stiff
polymers with a large bare persistence length, $\ell_B\ell_0/A^2\gg 1$, and sufficient
screening. In a systematic derivation, Barrat and Joanny found that the
electrostatic stiffening is actually scale-dependent and $\ell_e$ as given by
Eq.~(\ref{OSF}) is only realized at large scales~\cite{Barrat}.

The case of flexible polymers, however, remained somewhat unclear and a host of
controversial results can be found in the literature. On one hand, Khokhlov
and Khachaturian~\cite{KK} (KK) assumed that the OSF theory can be applied to a
chain of polyelectrolyte blobs in the case of weakly charged flexible
polyelectrolytes. At short length scales, the electrostatic repulsion is weaker
than the chain entropy. The electrostatic blob size, $\xi\simeq \ell_0 n^{1/2}$
where $n$ is the number of monomers (each of size $\ell_0$) in the blob, is
defined by the requirement that the electrostatic energy of the Gaussian
polymer coil is equal to the thermal energy, that is $(n\ell_0/A)^2\ell_B/
\xi\simeq 1$. The blob size follows as
\begin{equation}
\xi \simeq \ell_0\left(\frac{\ell_B\ell_0}{A^2}\right)^{-1/3} \label{xi}
\end{equation}
and the number of monomer in the blob is
\begin{equation}
n\simeq \left(\frac{\ell_B\ell_0}{A^2}\right)^{-2/3} \label{n}
\end{equation}
At a scaling level, the effective distance between charges along the direction
of the string of blobs is then renormalized by the linear density of monomers,
$A'^{-1}\simeq \frac{\ell_0 n}{\xi}A^{-1}$. By replacing $A$ by $A'$ in
Eq.~(\ref{OSF}), KK found
\begin{equation}
\ell_e=\ell_{\mathrm{KK}}\simeq
\left(\frac{\ell_B\ell_0}{A^2}\right)^{1/3}\frac{1}{\ell_0\kappa^2} \label{KK}
\end{equation}
which exhibits the same dependence in $\kappa^{-2}$ as the OSF result. This has
been confirmed by Li and Witten who studied the effect of chain fluctuations in
the presence of screening and showed that they do not affect the value of
$\ell_e$ found by KK~\cite{Li}.

On the other hand, using a variational approach by modelling the flexible
polyelectrolyte as a chain under tension, Barrat and
Joanny~\cite{Barrat} found a different $\kappa$ dependence:
\begin{equation}
\ell_e=\ell_{\mathrm{BJ}}\simeq\frac{1}{\kappa}
\label{BJ}
\end{equation}
This result is consistent with a scaling argument proposed by de~Gennes
\textit{et al.}~\cite{PGG} for solutions and has also been found by
Pfeuty~\cite{Pfeuty} using the Renormalization Group theory. More recent
renormalization calculations have been done by Liverpool and
Stapper~\cite{Liverpool2} and point to a similar (even sublinear) dependence of
the persistence length on $\kappa$. A number of different approaches led to
the same behaviour~: i) an approach based on minimization of an approximated free
energy~\cite{Schmidt}, ii) a $1/d$ expansion (where $d$ is the
space dimension)~\cite{Rudi}, and iii) variational approaches~\cite{Thirumalai1}
using the ``uniform expansion" method first derived by Edwards and
Singh~\cite{Singh,DE}.

It is important to mention that the result Eq.~(\ref{BJ}) poses problems at the
crossover charge $\ell_B\ell_0\simeq A^2$, between the Gaussian-persistent and
the persistent regime. The $\kappa$ dependence for $\ell_e$ is then not
continuous and a crossover formula should be found in theories. Among
theoretical works which predict Eq.~(\ref{BJ}), there is no satisfying answer to
this question.

Recently, Netz and Orland~\cite{Netz} used the most general Gaussian kernel for
the variational Hamiltonian~\cite{Jonsson1} and recovered the well-known three
distinct scaling regimes, defined according to the value of the screening
parameter $\kappa\ell_0$ and the charge parameter $\ell_B\ell_0/A^2$. For weakly
charged polymers and at large screening, one is in the \textit{Gaussian} regime
where the notion of electrostatic persistence length is meaningless. At small
screening, and according to the charge and the stiffness of the chain (for
$\ell_B\ell_0/A^2>1$), one reaches the persistent regime where $\ell_e$
is given by the OSF result, equation~(\ref{OSF}). The Gaussian-persistent regime
is reached when the electrostatic repulsion between monomers is weaker (for
$\ell_B\ell_0/A^2<1$) and the chain is crumpled at small length scales. It is
shown that in this last regime, the asymptotic electrostatic persistence length
(when $\kappa\ell_0\rightarrow 0$) is given by the KK result,
equation~(\ref{KK}).

Although the work by Netz and Orland is in accord with the growing consensus
that the KK result is asymptotically correct, it uses quite complicated
mathematics and cannot be easily extended to other systems, such as
polyelectrolyte solutions. Moreover, the persistence length is found quite
indirectly by matching the small-distance behaviour with the asymptotic swelling
range due to excluded volume interactions. Hence, the intermediate distance
range where the chain statistics is truly Gaussian at scales larger than the
persistence length is not considered. However, such an intermediate range always
exists provided that $\ell_p\gg \kappa^{-1}$ which is typically the case both in
the persistent and the Gaussian-persistent regimes~\cite{NetzRev}. The
persistence length should thus be determined by the crossover between the
persistent range and the intermediate Gaussian range. Finally, in order to
compare to experiments and simulations, it would be interesting to find the
$\kappa$-dependence also near the crossover, $\kappa\xi\simeq 1$, where
screening becomes important, and not only asymptotically. These are the reasons
that motivated this study where the variations of $\ell_e(\kappa)$ are
determined self-consistently by choosing $\ell_e$ as the variational parameter.

In a number of papers a similar single-parameter variational theory was used,
but the variational persistence length reflected the swelling behaviour of the
chain for large scales~\cite{Muthu2,Ghosh,Yethiraj,Donley1,Donley2,Winkler}.
This is borne out by the fact that the persistence length depends on the chain
length with a characteristic power law $\ell_e \sim N^\gamma$ which in the
infinite chain limit $N \rightarrow \infty$ gives the correct swelling behaviour
for the radius of gyration. In these calculations, the persistence length does
not reflect the \textit{mechanical} properties of the chain at small length
scales, because one would expect the persistence length to saturate to a finite
value as $N \rightarrow \infty$. In our paper, in contrast, we fix the
large-scale behaviour as being Gaussian (thus corresponding to the actual
intermediate regime for real chains) and the variational parameter reflects the
transition from the rod-like to the Gaussian behaviour. The reason for this
choice is that taking into account both the stiffening (at small length scales)
and the swelling (at large length scales) requires two distinct variational 
parameters, a formidable task which can be done only asymptotically and within
uncontrolled approximations~\cite{Netz}.


\section{Model}


In the following, we consider a Gaussian charged polymer
in a solution of monovalent salt (concentrations
$c_+$ for cations and $c_-$ for anions). The polymer chain of polymerization
index $N$ and a chain length $L=N\ell_0$ is supposed uniformly charged with
quenched charges $+e$. The distance between charges along the chain is given by
$A$. We remain at the linear level for electrostatic interactions, assuming a
Debye-H\"uckel screened interaction between charges due to the presence of
monovalent counterions and salt. As explained in the Introduction, we therefore
neglect non-linear effects connected to counterion condensation at low salt
concentrations (we plan to incorporate such effects in the future by using a
modified variational action; in the present paper we mostly try to resolve
inconsistencies in previous treatment of this linear model). We use a continuous
description of the polymer, parameterized by the curvilinear index $s$ and where
$\mathbf{r}(s)$ is the position of the monomer labelled by $s$.
The Hamiltonian of the system is~\cite{LNN,Muthu2,Winkler}
\begin{eqnarray}
\beta\mathcal{H}[\mathbf{r}] &=& \beta\mathcal{H}_{\mathrm{el}}[\mathbf{r}] +
\beta\mathcal{H}_{\mathrm{stiff}}[\mathbf{r}] +
\beta\mathcal{H}_{\mathrm{int}}[\mathbf{r}] \nonumber\\ &=& \frac12
\int_0^L\mathrm{d}s \left[\frac{3}{2\ell_0}\mathbf{\dot{r}}^2(s)+
\frac{3\ell_0}{2}\mathbf{\ddot{r}}^2(s)\right] \nonumber \\ &+&
\frac{\ell_B}{2A^2}\int_0^L \mathrm{d}s\int_0^L \mathrm{d}s'\frac{\exp(-\kappa
|\mathbf{r}(s)- \mathbf{r}(s')|)}{|\mathbf{r}(s)- \mathbf{r}(s')|} \label{H}
\end{eqnarray}
where $\beta=(k_BT)^{-1}$. The first term is the standard elasticity of entropic
origin, the second one the bending stiffness of the chain, and the last term
the electrostatic interaction. The partition function is
\begin{equation}
\mathcal{Z}=\int
\mathcal{D}\mathbf{r}\exp\{-\beta\mathcal{H}[\mathbf{r}]\}
\end{equation}
and is intractable due to the electrostatic term.

To make progress, we choose a variational Gaussian Hamiltonian which most
generally reads
\begin{equation}
\beta\mathcal{H}_0[\mathbf{r}]=
\frac{1}{2}\int_0^L \mathrm{d}s\int_0^L \mathrm{d}s'
\mathbf{r}(s)g^{-1}(s-s')\mathbf{r}(s') \label{Hvar}
\end{equation}
where $g^{-1}$ is the Gaussian kernel. Gaussian statistics implies that $g$ is
the monomer-monomer correlation function:
\begin{equation}
\langle\mathbf{r}(s)\mathbf{r}(s')\rangle_0 = 3g(s-s')
\end{equation}
where $\langle ... \rangle_0$ denotes the expectation value computed with the
variational Hamiltonian~(\ref{Hvar}). It is related to the squared monomer
separation as
\begin{equation}
\langle[\mathbf{r}(s)-\mathbf{r}(s')]^2\rangle_0
= G(s-s') = 6g(0)-6g(s-s')
\label{Gdef}
\end{equation}
The tangent-tangent correlation function is simply
\begin{equation}
\langle\mathbf{\dot{r}}(s)\mathbf{\dot{r}}(s')\rangle_0 =
3\frac{\partial^2}{\partial s \partial
s'}g(s-s')=\frac12 G''(s-s')
\label{ttcorr}
\end{equation}

The variational free energy reads in the Gibbs-Bogoliubov form
\begin{equation}
F_{\mathrm{var}}=F_0 + \langle\mathcal{H}-\mathcal{H}_0 \rangle_0
\label{Fvar}
\end{equation}
where $\beta F_0=-\ln\mathcal{Z}_0$ is the free energy associated
with the variational Hamiltonian defined in equation~(\ref{Hvar}).
The Gibbs inequality ensures that $F_{\mathrm{var}}\geq
F_{\mathrm{exact}}$ when $F_{\mathrm{var}}$ is minimized
with respect to the variational parameters. As we are interested in the
persistence length, we will choose the electrostatic contribution $\ell_e$  (to
be defined below) as the only variational parameter. The value of the
persistence length will thus be determined by minimizing equation~(\ref{Fvar})
with respect to $\ell_e$. It must be emphasized, however, that we will restrict
our choice of variational Hamiltonians to the subclass of non-swollen (ideal)
correlations at large scales, and by  doing so, we do not obtain the true
minimum of the variational free energy.

Before starting explicit calculations, it is useful to consider
the Hamiltonian~(\ref{H}) and to rescale all lengths by the bare
persistence length according to $r=\ell_0\tilde{r}$ and
$s=\ell_0\tilde{s}$. We obtain
\begin{eqnarray}
\beta\mathcal{H}[\mathbf{\tilde{r}}] &=&
\frac34 \int_0^N\mathrm{d}\tilde{s}
\left[\mathbf{\dot{\tilde{r}}}^2(\tilde{s})+
\mathbf{\ddot{\tilde{r}}}^2(\tilde{s})\right] \label{Hr}  \\ &+&
\frac12\frac{\ell_B\ell_0}{A^2}\int_0^N \mathrm{d}\tilde{s}\int_0^N
\mathrm{d}\tilde{s}'\frac{\exp(-\kappa\ell_0 |\mathbf{\tilde{r}}(\tilde{s})-
\mathbf{\tilde{r}}(\tilde{s}')|)}{|\mathbf{\tilde{r}}(\tilde{s})-
\mathbf{\tilde{r}}(\tilde{s}')|} \nonumber
\end{eqnarray}
with three dimensionless parameters, namely the charge parameter,
$\ell_B\ell_0/A^2$, the screening parameter $\kappa\ell_0$ and the
polymerization index $N$. In this paper, we consider the thermodynamic
limit $N\to\infty$ and therefore are only left with two parameters. 

In the limit $\ell_B\ell_0/A^2=0$ the tangent-tangent
correlation function can be calculated exactly  
from Hamiltonian~(\ref{H}) and reads
\begin{equation}
\langle\mathbf{\dot{r}}(s)\mathbf{\dot{r}}(0)\rangle_{\mathrm{wlc}} =
\exp\left(-\frac{|s|}{\ell_0}\right)
\label{WLC1}
\end{equation}
while  the squared monomer separation is given by
\begin{equation}
G_{\mathrm{wlc}}(s)=2\ell_0^2\left[\frac{|s|}{\ell_0}-1 +
\exp\left(-\frac{|s|}{\ell_0}\right)\right] \label{WLC2}
\end{equation}
The correlation function therefore is identical to the worm-like-chain (WLC) model
which has been first proposed by Kratky and Porod for describing neutral semiflexible
chains~\cite{desCloiseaux}.

In our variational calculation, we will choose a modified WLC Hamiltonian, or,
equivalently, a modified correlation function $G(s)$, which will reflect
non-zero values of the charge parameter $\ell_B\ell_0/A^2$. The choice of the
variational correlation function will contain our knowledge of the behaviour of
charged chains. For weakly charged and flexible chains, the electrostatic
repulsion between two adjacent charged segments of length $\ell_0$ and charge
$\ell_0/A$ is weaker than the thermal energy, $k_BT$, for
$(\ell_0/A)^2\ell_B/\ell_0<1$, and thus is not sufficient to align the two
segments. This threshold is equivalent to the condition
\begin{equation}
\frac{\ell_B\ell_0}{A^2}<1
\end{equation}
In this case, the statistics of the chain is Gaussian at small scales and the chain has
locally a crumpled configuration. The electrostatic blob size~\cite{PGG}, $\xi$, is
thus defined at the scaling level by Eq.~(\ref{xi}). This blob size is found
using a scaling argument but can also be found variationally as shown in
Refs.~\cite{PGG,Barrat} plus a logarithmic correction~\cite{Netz}. In the
following, we neglect logarithmic corrections and use Eq.~(\ref{n}) for $n$ in
the rest of the paper. For weakly charged chains, we choose a variational
correlation function in accordance with these scaling results and which exhibits
the crumpled configuration at small scales (Section~4). The case of highly
charged and stiff chains for $\ell_B\ell_0/A^2>1$, which is treated in
Section~5, is more subtle since here the variational correlation function has to
be treated in a way such that all integrals occurring in the variational free
energy are regular in the small distance limit.


\section{Flexible polyelectrolytes}


At scales smaller than the electrostatic blob size ($\tilde{s}<n$) the
chain conformation is Gaussian, as argued above. For larger scales
($\tilde{s}>n$), the chain shows, for sufficiently large charge parameter,
electrostatic stiffness and should be described by an analogue of
Eqs.~(\ref{WLC1}) and~(\ref{WLC2}), provided that we properly rescale the
contour length by the blob size.

To simplify the mathematics, we in this section use a flexible polymer model,
\begin{eqnarray}
\beta\mathcal{H}[\mathbf{r}] &=& \frac{3}{2\ell_0} \int_0^L\mathrm{d}s
\mathbf{\dot{r}}^2(s)  \nonumber \\ &+&
\frac{\ell_B}{2A^2}\int_0^L \mathrm{d}s\int_0^L \mathrm{d}s'\frac{\exp(-\kappa
|\mathbf{r}(s)- \mathbf{r}(s')|)}{|\mathbf{r}(s)- \mathbf{r}(s')|} \label{Hf}
\end{eqnarray}
The  simplest variational tangent-tangent correlation function can be written as
\begin{equation}
\langle\mathbf{\dot{r}}(s)\mathbf{\dot{r}}(0)\rangle_0 = \frac12 G''(s)=
\ell_0\delta(s) + \frac{1}{n}\exp\left(-\frac{|s|}{n^{1/2}\ell_e}\right)
\label{WLC3}
\end{equation}
which reproduces the crumpled statistics following from Eq.~(\ref{Hf}) at small
scales, and also a persistent behaviour with a persistence length $\ell_e$ (or
$n^{1/2}\ell_e$ along the contour of the chain) at large scales. 
The factor $n^{-1}$ in front of
the second term ensures the correct crossover between the Gaussian and rod-like
behaviour. By integrating twice, we obtain the mean-squared monomer-monomer
separation \begin{equation} G(s)=\ell_0|s|+
2\ell_e^2\left[\frac{|s|}{n^{1/2}\ell_e}-1
+\exp\left(-\frac{|s|}{n^{1/2}\ell_e}\right)\right] \label{WLC4}
\end{equation}
We remark that our variational choice is exactly the same as the one proposed by
Barrat and Joanny~\cite{Barrat}. This can be shown explicitly by integrating out
the fluctuating tension $\mathbf{t}(s)$ (which can be done exactly) in their
partition function.

In the following, we compute the different terms of the variational
free energy equation~(\ref{Fvar})
using equations~(\ref{H}),~(\ref{Hvar}) and~(\ref{WLC3}). We denote by
$h(\omega)$ the Fourier transform of the tangent-tangent correlation function
\begin{eqnarray}
h(\omega) &=& \int_{-\infty}^{\infty}\mathrm{d}s
\langle\mathbf{\dot{r}}(s)\mathbf{\dot{r}}(0)\rangle_0 \mathrm{e}^{\iota \omega
s} \nonumber\\ &=& \int_0^{\infty}\mathrm{d}s G''(s)\cos(\omega s)
\label{hdef}
\end{eqnarray}
where we used that  $G''$ is an even function. We find
\begin{equation}
h(\omega) = \ell_0\left[1+\frac{\ell_e}{\xi}
\frac{2}{1+n\ell_e^2\omega^2}\right] \label{hw}
\end{equation}
The average value of the elastic part is easily found to be
\begin{eqnarray}
\beta\langle \mathcal{H}_{\mathrm{el}}\rangle_0 &=&
\int_{-\infty}^{\infty} \frac{\mathrm{d}\omega}{2\pi}\frac{3L}{2\ell_0}h(\omega)
= \frac{3L}{2\ell_0} \langle\mathbf{\dot{r}}(0)\mathbf{\dot{r}}(0)\rangle_0
\nonumber\\ &=&\frac{3N}{2}
\int_{-\infty}^{\infty}\frac{\mathrm{d}\omega}{2\pi}+\frac32\frac{N}{n}
\label{Hel}
\end{eqnarray}
which does not depend on the variational parameter. Furthermore, it is easy to
check that $\langle\mathcal{H}_0\rangle_0$ is a constant independent of
$\ell_e$. The entropy term is given by
\begin{eqnarray}
\beta F_0 &=& -\ln\mathcal{Z}_0 \nonumber\\ &=&
-\frac{3L}{2}\int_{-\infty}^{\infty}\frac{\mathrm{d}\omega}{2\pi} \ln\left[
\frac{h(\omega)}{\ell_0^3\omega^2}\right]
\end{eqnarray}
By substracting the free energy part associated with the ultraviolet divergence
($\omega\to \infty$), i.e. the small length-scale fluctuations which do not
depend on the variational parameter,
\begin{equation}
\beta F_{\mathrm{uv}} = \frac{3L}{2}
\int_{-\infty}^{\infty}\frac{\mathrm{d}\omega}{2\pi}\ln(\ell_0^2\omega^2)
\label{Fuv}
\end{equation}
we find
\begin{eqnarray}
\beta \left(F_0-F_{\mathrm{uv}}\right) &=&
-\frac32 \frac{N}{n} \frac{\xi}{\ell_e} \int_{-\infty}^{\infty}
\frac{\mathrm{d}x}{2\pi}
\ln\left(1+\frac{\ell_e}{\xi}\frac{2}{1+x^2}\right)\nonumber\\ &=&\frac32
\frac{N}{n}\frac{\xi}{\ell_e}\left(1-\sqrt{1+2\frac{\ell_e}{\xi}}\right)
\label{entropymwlc}
\end{eqnarray}
In the limit $N\rightarrow \infty$, the interaction term reduces to
\begin{eqnarray}
\beta \langle \mathcal{H}_{\mathrm{int}} \rangle_0 &=& \frac{L\ell_B}{A^2}
\int_0^{\infty} \mathrm{d}s\left\langle\frac{\exp(-\kappa
|\mathbf{r}(s)- \mathbf{r}(0)|)}{|\mathbf{r}(s)-
\mathbf{r}(0)|}\right\rangle_0 \label{Hint2}\\
&=& \frac{L\ell_B}{A^2} \int_0^{\infty} \mathrm{d}s
\int\frac{\mathrm{d}^3\mathbf{q}}{(2\pi)^3}
\frac{4\pi}{\kappa^2+\mathbf{q}^2} \left\langle
\mathrm{e}^{\imath\mathbf{q}|\mathbf{r}(s)-\mathbf{r}(0)|}\right\rangle_0
\nonumber\\ &=& \frac{L\ell_B}{A^2} \frac{2}{\pi}\int_0^{\infty}\mathrm{d}s
\int_0^{\infty}\mathrm{d}q\frac{q^2}{\kappa^2+q^2}\mathrm{e}^{-\frac16
q^2G(s)} \nonumber
\end{eqnarray}
By inserting equation~(\ref{WLC3}) in the last equation, using the definition of
$n$ and after a few algebraic calculations we obtain
\begin{eqnarray}
\frac{n}{N}\beta\langle\mathcal{H}_{\mathrm{int}}\rangle_{0}
&=&\int_0^{\infty} \mathrm{d}x \left\{\sqrt{\frac{6}{\pi}}
\frac{1}{\sqrt{G(x)}}\right.\label{Hint}\\
&-&\left.\kappa\ell_e\mathrm{e}^{\frac16(\kappa\ell_e)^2G(x)}
\mathrm{erfc}\left[\frac{\kappa\ell_e}{\sqrt{6}}\sqrt{G(x)}\right]\right\}
\nonumber
\end{eqnarray}
where
\begin{equation}
G(x) = x\frac{\xi}{\ell_e} + 2(\mathrm{e}^{-x}-1+x)
\end{equation}
and the complementary error function is
\begin{equation}
\mathrm{erfc}(x)=\frac{2}{\sqrt{\pi}}\int_x^{\infty}\mathrm{d}t\exp(-t^2)
\end{equation}
At scales larger than the electrostatic blob size $\xi$, tangents are
correlated over distances up to the persistence length $\ell_e$ if the Debye
length, $\kappa^{-1}$, is larger than $\xi$, that is, if $\kappa\xi<1$. The case
where $\kappa\xi>1$ and $n>1$ corresponds to the Gaussian regime~\cite{Netz} and
the notion of electrostatic persistence length is then meaningless. We have
checked that, indeed, $\ell_e$ goes to 0 for $\kappa\xi>1$.

\begin{figure}[htb]
\begin{center}
\includegraphics[height=6cm]{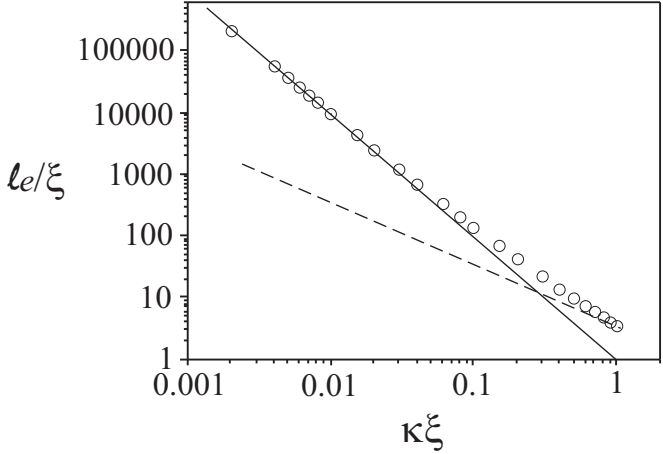}
\end{center}
\caption{Numerical minimization of the variational equation~(\ref{Fvar}) with
respect to $\ell_e$: log-log plot of the renormalized persistence length,
$\ell_e/\xi$, vs. the renormalized Debye parameter, $\kappa\xi$ (where $\xi$ is
the electrostatic blob size). The solid line denotes the function $\ell_e =
\frac{1}{\xi\kappa^2}$ while the broken line is proportional to $\kappa^{-1}$.}
\label{minim}
\end{figure}

The minimization of the variational free energy per blob, $n \beta
F_{\mathrm{var}}/N$, with respect to $\ell_e$ is first done numerically. The
result $\ell_e(\kappa)$ for $\kappa\xi$ varying from 0.001 to 1 is shown in
Figure~\ref{minim} in a log-log plot. We observe that the electrostatic
persistence length decreases with $\kappa$ following the power law $\ell_e \sim
\kappa^{-2}$ in the limit of small $\kappa\xi$ ($\kappa\xi<0.1$). The solid line
denotes the function
\begin{equation}
\ell_e = \frac{1}{\xi\kappa^2}
\label{result}
\end{equation}
which demonstrates that the result found by Khokhlov and Khachaturian by scaling
arguments is correct for Debye lengths larger than 10 blob sizes. For smaller
Debye lengths ($0.1<\kappa\xi<1$), our results show a smooth crossover to the
results obtained in Ref.~\cite{Barrat}, Eq.~(\ref{BJ}), shown in
Figure~\ref{minim} as a broken line. It should be noted that this part of the
curve ($0.1<\kappa\xi<1$) varies depending on the variational choice and is not
universal (see Appendix). To understand where the discrepancy between our
asymptotic scaling and the result by Barrat and Joanny comes from, we analyze
our variational free energy in the limit of weak screening.

From Eq.~(\ref{entropymwlc}), we can approximate the entropy per
electrostatic blob by
\begin{equation} \label{eq32}
\frac{n}{N}\beta(F_0-F_{\mathrm{uv}})\simeq
-\frac{3}{\sqrt2}\left(\frac{\xi}{\ell_e}\right)^{1/2}
\end{equation}
in the limit $\ell_e/\xi\rightarrow\infty$, which is the low-screening limit.
Obviously, this limiting formula is different from the result found by Barrat
and Joanny for the entropy, $\frac{n}{N}\beta F_0\simeq\ln(\ell_e/\xi)$ [see
Eq.~(7) of Ref.~\cite{Barrat}]. This different dependence of the entropy on
$\ell_e$ is the sole reason for their different scaling of the persistence
length, resulting in $\ell_e\sim\kappa^{-1}$.

\begin{figure}[htb]
\begin{center}
\begin{tabular}{cc}
\includegraphics[height=4.7cm]{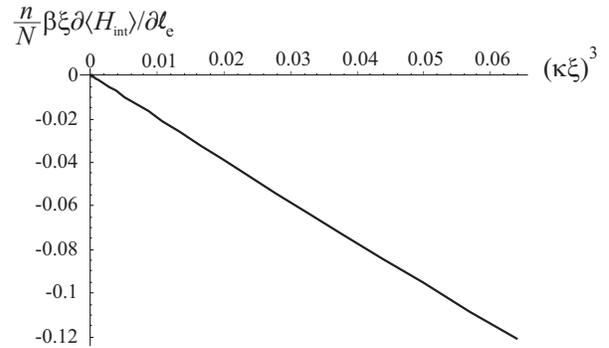}
\end{tabular}
\end{center}
\caption{Numerical check of Eq.~(\ref{dHint}). The first derivative of the
interaction term per electrostatic blob, $\frac{n}{N}\beta\xi\partial\langle
\mathcal{H}_{\mathrm{int}}
\rangle_0/\partial\ell_e\vert_{\ell_e=(\kappa^2\xi)^{-1}}$, decreases as
$-(\kappa\xi)^3$ when $\ell_e$ is assumed to follow the KK law,
$\ell_e\simeq(\kappa^2\xi)^{-1}$.} \label{Hintfig}
\end{figure}

The interaction term  equation~(\ref{Hint}) is more difficult to evaluate. The
two terms cannot be computed separately because of a cancellation of
divergences. To make analytical progress, we assume the
result~(\ref{result}) to be  valid and replace $\ell_e$ by
$1/(\xi\kappa^{2})$ in the first two terms of equation~(\ref{Hint}). We then
look at the limit $\kappa\xi\rightarrow 0$. We find numerically for the leading
term
\begin{equation}
\frac{n}{N}\beta\langle \mathcal{H}_{\mathrm{int}}
\rangle_0 \simeq -\ln(\kappa\xi)
\end{equation}
It corresponds to the electrostatic energy of the blob in a cylindrical
geometry and diverges logarithmically when $\kappa$ goes to 0. This term
is independent of the electrostatic persistent length. To find the next leading
term, we calculate the asymptotic behavior of the derivative of
$\frac{n}{N}\beta \langle \mathcal{H}_{\mathrm{int}} (\ell_e,\kappa)\rangle_0$
with respect to $\ell_e$ and then assume the KK law. As is visualized  in
Figure~\ref{Hintfig}, we find
\begin{equation}
\frac{n}{N}\beta\xi\left.\frac{\partial\langle
\mathcal{H}_{\mathrm{int}}\rangle_0}{\partial\ell_e}
\right\vert_{\ell_e=(\kappa^2\xi)^{-1}} \sim -(\kappa\xi)^3 +
\mathcal{O}((\kappa\xi)^5) \label{dHint}
\end{equation}
Keeping the first terms of the asymptotic expansion of $\partial
F_{\mathrm{var}}/\partial\ell_e$, we thus get
\begin{equation}
\frac{\partial F_0}{\partial\ell_e}+\left.\frac{\partial\langle
\mathcal{H}_{\mathrm{int}}\rangle_0}{\partial\ell_e}
\right\vert_{\ell_e=(\kappa^2\xi)^{-1}} \sim \frac{\xi^{1/2}}{\ell_e^{3/2}}
-\kappa^3\xi^2 =0
\end{equation}
and thus recover Eq.~(\ref{result}) in the asymptotic limit
$\kappa\xi\rightarrow 0$ (within a numerical coefficient of the order of unity)
in a self-consistent way.

Besides this asymptotic but cumbersome way, it is also possible to find the
leading term in $(\kappa\ell_e)^{-1}$ for the interaction term by using an
approximated structure factor for the semi-flexible chain~\cite{Barrat}.
Starting from Eq.~(\ref{Hint2}), we see that
\begin{eqnarray}
\beta \langle \mathcal{H}_{\mathrm{int}} \rangle_0 &=& \frac{L\ell_B}{A^2}
\int_0^{\infty} \mathrm{d}s
\int\frac{\mathrm{d}^3\mathbf{q}}{(2\pi)^3}\frac{4\pi}{\kappa^2+\mathbf{q}^2}
\left\langle
\mathrm{e}^{\imath\mathbf{q}|\mathbf{r}(s)-\mathbf{r}(0)|}\right\rangle_0
\nonumber \\ &=& \frac{L\ell_B}{2A^2}\int\frac{\mathrm{d}^3\mathbf{q}}{(2\pi)^3}
\frac{4\pi}{\kappa^2+q^2} S_0^{\infty}(q) \label{HintS}
\end{eqnarray}
where the structure factor, $S_0^{\infty}(q)$, of an infinite chain computed
using the variational Hamiltonian, $\mathcal{H}_0$, is given by
\begin{equation}
S_0^{\infty}(q)=2\int_0^{\infty} \mathrm{d}s \left\langle
\mathrm{e}^{\imath\mathbf{q}|\mathbf{r}(s)-\mathbf{r}(0)|}\right\rangle_0
\end{equation}
In the limit $\ell_e/\xi \rightarrow\infty$, the chain can be viewed as a
semi-flexible chain made up of $N/n$ rods of length $\xi$ and of persistence
length $\ell_e$. We thus neglect the internal blob structure. We can use an
interpolating formula for the polyelectrolyte structure factor,
$S_0^{\infty}(q)=(1+q\ell_e)/(q^2\ell_e)$ which exhibits the correct limiting
behaviour for small momenta in $1/(q^2\ell_e)$ and for large momenta in
$1/q$~\cite{desCloiseaux}. Using a upper cutoff at $q\simeq\xi^{-1}$ in
Eq.~(\ref{HintS}), we easily find 
\begin{equation}
\frac{n}{N}\beta \langle\mathcal{H}_{\mathrm{int}}\rangle_0
\simeq -\frac12 \ln(\kappa\xi) + \frac{\pi}{4} \frac{1}{\kappa\ell_e} +
\mathcal{O}\left(\frac{1}{(\kappa\ell_e)^{2}}\right) \label{expa}
\end{equation}
Taking a  derivative with respect to $\ell_e$ and inserting the KK
scaling Eq.(\ref{result}), we recover Eq.~(\ref{dHint}). Hence, we note that 
the $\kappa^{-2}$ dependence of $\ell_e$ comes directly from the balance between
the entropy of the chain in equation (\ref{eq32})
 and the electrostatic energy
in equation (\ref{expa})  in the limit of weak screening.

\section{Stiff polyelectrolytes}

\begin{figure}[htb]
\begin{center}
\includegraphics[height=4.5cm]{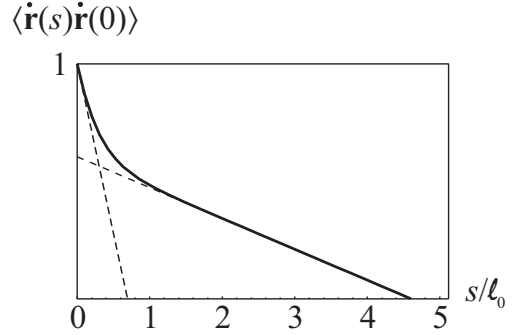}
\end{center}
\caption{Log plot of the variational tangent-tangent correlation function vs.
$\tilde{s}=s/\ell_0$ [see Eq.~(\ref{defgOSF})]. The broken lines are the two
asymptotic behaviours given by Eq.~(\ref{BJb}). The parameter values are
$\kappa\ell_0=1$ and $\ellt_e=10$} \label{gOSF}
\end{figure}

In the \textit{persistent} regime, for stiff and strongly charged polymers,
$\ell_0\ell_B/A^2>1$, and at small screening,
$\ell_0\ell_B/A^2>(\ell_0\kappa)^2$, the above variational kernel must be
modified. According to OSF theory, which is based on the electrostatic
contribution to the energy of a uniformly bent rod, the electrostatic
monomer-monomer interactions are already relevant on length scales comparable
to the bare persistent length, leading to an effective persistent length given
by equation~(\ref{OSF}). However, Barrat and Joanny have shown within a
quadratic perturbation treatment that the persistence length is in fact
scale-dependent and that the OSF prediction corresponds to the large-scale
limit~\cite{Barrat}. Since all these calculations are perturbative and only
valid for weakly bent chains, it is interesting to have a variational
determination of the effective persistent length $\ell_p=\ell_0+\ell_e$ in this
persistent regime. The first choice would be the WLC model with a variable
persistence length. Treating the persistence length $\ell_p$ as a variational
parameter, the tangent-tangent correlation function and the squared monomer
separation are given by Eqs.~(\ref{WLC1}) and~(\ref{WLC2}) where $\ell_0$ is
replaced by $\ell_p$. For an \textit{infinite} chain, $N\rightarrow\infty$, a
Gaussian Hamiltonian which leads to these equations is Eq.~(\ref{H}) with
$\ell_B/A^2=0$ and $\ell_p$ instead of $\ell_0$. However, this variational
Hamiltonian with $\ell_p$ as the variational parameter does not work because the
variational entropic free energy diverges at small length scales
\begin{equation}
\beta F_{\mathrm{wlc}}=-\frac{3L}{2}\int_{-\infty}^{\infty}
\frac{\mathrm{d}\omega}{2\pi}\ln\left[\frac{\ell_p}{\ell_0}
\frac{2}{\ell_0^2\omega^2(1+\ell_p^2\omega^2)}\right]
\end{equation}
since the integrand behaves like $\ln(\ell_0\ell_p\omega^4)$ when
$\omega\rightarrow\infty$. This divergence depends on the variational parameter,
and therefore, a variational approach based on such a Hamiltonian is
ill-defined. This fact is reflected by Monte-Carlo simulations and perturbative
calculations for the tangent-tangent correlation function of a charged chain~:
at small length scales, the decay of this correlation function is given by the
bare (mechanical) persistence length, and only at larger length scales one finds
a crossover to the electrostatic persistence~\cite{Barrat}. This can be
described by a scale-dependent persistence length
$\ell_p(\omega)=\ell_0+\ell_e(\omega)$ where $\ell_e(\omega)\simeq 0$ for
$\omega\rightarrow\infty$ and $\ell_e(\omega)\simeq \ell_{\mathrm{OSF}}$ for
$\omega\rightarrow 0$ which leads to a tangent-tangent correlation with the
limiting behaviours
\begin{equation}
\langle\mathbf{\dot{r}}(s)\mathbf{\dot{r}}(0)\rangle \simeq
\left\{\begin{array}{lcr}
1-\frac{|s|}{\ell_0} & \mathrm{for} & |s|<s_c \\
\exp\left(-\frac{|s|}{\ell_0+\ell_{\mathrm{OSF}}}\right) & \mathrm{for} &
|s|>s_c \label{BJb}
\end{array}\right.
\end{equation}
where the crossover contour length $s_c$ has been determined in
Ref.~\cite{Barrat} and is
\begin{equation}
s_c\simeq \frac{1}{\kappa\sqrt{1+\ell_{\mathrm{OSF}}/\ell_0}}
\label{sc}
\end{equation}
It transpires that the variational correlation function has to be chosen in
accord with these limits. A simple and continuous choice for the tangent-tangent
correlation function which fulfills Eq.~(\ref{BJb}) is
\begin{eqnarray}
\langle\mathbf{\dot{r}}(s)\mathbf{\dot{r}}(0)\rangle_0 &=&
B\exp\left(-\frac{|s|}{\ell_e+\ell_0}\right)
\label{defgOSF}\\ &+& (1-B)\exp\left[-|s|
\frac{\ell_e+(1-B)\ell_0}{\ell_0(\ell_e+\ell_0)(1-B)}\right]
\nonumber
\end{eqnarray}
where $\ell_e$ is the variational parameter and the factor $B$ is chosen such as
to satisfy continuity
\begin{equation}
B=\exp\left[-\frac{\ell_e}{\kappa\sqrt{\ell_0}(\ell_e+\ell_0)^{3/
2}}\right]
\end{equation}
Hence, the crossover $s_c=\sqrt{\ell_0}/(\kappa\sqrt{\ell_e+\ell_0})$ depends on
the variational parameter. The case of a neutral semi-flexible polymer is
recovered by taking $s_c\to\infty$ (which implies $B=0$) in Eq.~(\ref{defgOSF}).
The tangent-tangent correlation function is plotted in Figure~\ref{gOSF} for
$\kappa\ell_0=1$ and $\ell_e=10\ell_0$.

The Fourier transform, defined in Eq.~(\ref{hdef}), is easily computed
\begin{eqnarray}
h(\omega) &=& 2B\frac{\ell_e+\ell_0}{1+(\ell_e+\ell_0)^2\omega^2}\\&+&
\frac{2(1-B)^2(\ell_e+\ell_0)}{\ell_e+(1-B)\ell_0}
\frac{\ell_0}{1+\left[\frac{(\ell_e+\ell_0)(1-B)\ell_0}{\ell_e+(1-B)\ell_0}
\right]^2 \omega^2} \nonumber
\end{eqnarray}
Following the lines of Section~3, we obtain for the entropy
\begin{eqnarray}
\beta(F_0-F_{\mathrm{n}}) &=&
-\frac{3L}{2}\int_{-\infty}^{\infty}\frac{\mathrm{d}\omega}{2\pi}
\ln\left[\frac{1}{2\ell_0}(1+\ell_0^2\omega^2) h(\omega)\right] \nonumber\\ &=&
-\frac{3N}{2(\ellt_e+1)}\left[\ellt_e+1-\frac{\ellt_e}{1-B}\right.
\label{entropystiff} \\ &+& \left.
\frac{\sqrt{(1-B+\ellt_e)(1-B+B\ellt_e)}}{\sqrt{\ellt_e+1}(1-B)}\right]
\nonumber
\end{eqnarray}
where $\ellt_e$ is the adimensional electrostatic persistence length
\begin{equation}
\ellt_e=\frac{\ell_e}{\ell_0}
\end{equation}
and
\begin{equation}
\beta F_{\mathrm{n}}=\frac{3L}{2}\int_{-\infty}^{\infty}
\frac{\mathrm{d}\omega}{2\pi}\ln\left[\frac12
\omega^2\ell_0^2(1+\ell_0^2\omega^2)\right]
\end{equation}
is the entropy of a neutral semi-flexible polymer. The elastic term is
\begin{eqnarray}
\beta\langle \mathcal{H}_{\mathrm{el}}\rangle_0 &=& \frac{3L}{4}
\left(\frac{1}{\ell_0} -\ell_0 \frac{\mathrm{d}^2}{\mathrm{d}s^2}\right)
\langle\mathbf{\dot{r}}(0)\mathbf{\dot{r}}(0)\rangle_0 \nonumber\\
&=& -\frac{3N}{4} \frac{B\ellt_e^2}{(1-B)(\ellt_e+1)^2} \label{Helstiff}
\end{eqnarray}
The electrostatic energy is found using Eq.~(\ref{Hint2}) which yields
\begin{eqnarray}
\beta\langle\mathcal{H}_{\mathrm{int}}\rangle_{0}&=&
N\frac{\ell_B\ell_0}{A^2}\int_0^{\infty} \mathrm{d}x \left\{\sqrt{\frac{3}{\pi}}
\frac{1}{\sqrt{f(x)}} -\kappa(\ellt_e+1)
\right. \nonumber\\
&\times&\left.\mathrm{erfc}\left[\frac{\kappa(\ellt_e+1)}{\sqrt{3}}
\sqrt{f(x)}\right]
\mathrm{e}^{\frac13\kappa^2(\ellt_e+1)^2f(x)}\right\} \label{Hintstiff}
\end{eqnarray}
where
\begin{eqnarray}
f(x) &=& B(\mathrm{e}^{-x}-1+x)+\frac{(1-B)^3}{(\ellt_e+1-B)^2}\\ &\times&
\left(\mathrm{e}^{-x\frac{\ellt_e+1-B}{1-B}}+x\frac{\ellt_e+1-B}{1-B}-1
\right) \nonumber
\end{eqnarray}
The interaction term, Eq.~(\ref{Hintstiff}), diverges
logarithmically for small $x$ (rod-like divergence). In the
computation, we thus subtract $N
\frac{\ell_B\ell_0}{A^2}\sqrt{\frac{6}{\pi}}\int_0^{\infty}\frac{\mathrm{d}x}{x(
1+x^3)}$ which does not depend on the variational parameter.

\begin{figure}[htb]
\begin{center}
\includegraphics[height=6cm]{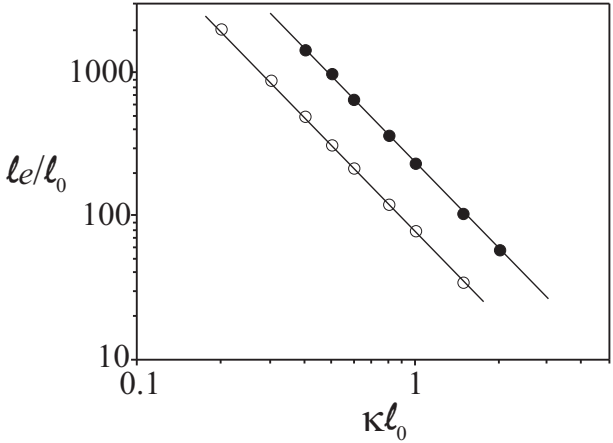}
\end{center}
\caption{Numerical minimization of the variational equation free energy with
respect to $\ell_e$ for stiff chains~: log-log plot of the renormalized
persistence length, $\ell_e/\ell_0$, vs. the renormalized Debye parameter,
$\kappa\ell_0$ for two values of $\ell_b\ell_0/A^2$~: 2.25 (open circles) and 4
(full circles). Solid lines are the power law $\kappa^{-2}$.}
\label{minimizationOSF}
\end{figure}

The minimization of the variational free energy is done numerically and
the result is plotted in Figure~\ref{minimizationOSF} for
$\ell_B\ell_0/A^2=2.25$ and 4. We observe that the OSF result is recovered with
a decrease $\ellt_e\sim\kappa^{-2}$ in both cases.

We should add that we find the persistence length to depend on the charge
parameter $\ell_B\ell_0/A^2$ as $\ellt_e\sim(\ell_B\ell_0/A^2)^2$ which is
different from the linear dependence found by OSF [Eq.~(\ref{OSF})]. At this
point we have no clear interpretation for this discrepancy. Because of numerical
limitations, we could calculate the effective persistence length only for
moderate values of the charge parameter up to $\ell_B\ell_0/A^2 = 8$, which is
the range  of parameters found for fully charged flexible (synthetic) polymers.
This parameter range is thus very far from the value $\ell_B\ell_0/A^2 = 150$
used in the simulations by Barrat and Joanny. An explanation could be that close
to the crossover $\ell_B\ell_0/A^2=1$, the correction, $\varphi$, to the OSF law
according to
\begin{equation}
\ellt_e=\frac14
\frac{\ell_B\ell_0}{A^2}\frac{1}{(\kappa\ell_0)^{2}}\,
\varphi\left(\kappa\ell_0,
\frac{\ell_B\ell_0}{A^2}\right) \quad\mathrm{with}\quad \varphi(0,\infty)=1
\end{equation}
is very large. This point will be studied by suitable simulations in the future.


\section{Discussion}


In this section, we discuss the results of Section~3 and compare our approach to
those of Barrat and Joanny~\cite{Barrat}, Schmidt~\cite{Schmidt}, Ha and
Thirumalai~\cite{Thirumalai1,Thirumalai2} and Muthukumar \textit{et
al.}~\cite{Muthu2,Ghosh}. All these works are variational approaches with only
one variational parameter. However, as explained in the Introduction, we
shall distinguish between two types of works, according to the physical
interpretation of this parameter. In Ref.~\cite{Muthu2,Ghosh}, it reflects the
swelling behaviour at large length scales and reproduce the correct swelling law
as a function of $N$. This variational parameter is then not interpreted as the
true mechanical persistence length. In the other
references~\cite{Schmidt,Barrat,Thirumalai1,Thirumalai2}, the variational
parameter is interpreted as a persistence length and the swelling behaviour at
large scales is not considered. Our variational calculation belongs to this
category. We show that, in this case, the persistence length is finite for an
infinite chain, $N\to\infty$.

As mentioned in Section~3, our variational calculation is similar to
Barrat and Joanny calculation. They find the result Eq.~(\ref{expa}) for the
interaction term, by approximating the structure factor of a polyelectrolyte and
by using an upper cutoff at $q\simeq\xi^{-1}$. Hence they ``smooth"
electrostatic interactions inside the electrostatic blobs. This approximation is
valid \textit{only} in the limit $\ell_e/\xi\to\infty$, i.e. in the limit of
weak screening. However, using the same cutoff at $q\simeq\xi^{-1}$ in the
calculation of the entropic term, they find $\beta F_0\simeq
\frac{N}{n}\ln(\ell_e/\xi)$, which is different from our result,
Eq.~(\ref{entropymwlc}), and leads to their law $\ell_e\simeq\kappa^{-1}$.
Evidently, this un-controlled cutoff leads to the correct result for the
interaction term but not for the entropy.

Schmidt~\cite{Schmidt} use a different variational formulation, where the
elastic energy is calculated for a stretched random flight chain and the
electrostatic energy is approximated using an hybrid rod-coil model. This leads
to a weaker dependence than $\kappa^{-2}$ for $\ell_e$ but which does not follow
a power law and tends to vary as $\kappa^{-1}$ at high ionic strength. However,
$\ell_e$ increases with the contour length and seems to diverge for
$N\to\infty$. Moreover, as carefully explained in Ref.~\cite{Schmidt}, this
calculation does not apply to the low ionic strength limit in view of the
approximations made.

Ha and Thirumalai~\cite{Thirumalai1,Thirumalai2} use the same variational
squared monomer separation, Eq.~(\ref{WLC4}), but follow the uniform expansion
method, first developed by Edwards and Singh~\cite{Singh,DE}. This perturbative
variational method is performed on the mean-square end-to-end distance, $\langle
R^2\rangle$. It consists in calculating self-consistently the persistence
length by requiring $\langle R^2\rangle -\langle R^2\rangle_0$ to be zero at
first order in $\ell_0\ell_B/A^2$. This method has proved to be powerful in the
calculation of the Flory exponent for the end-to-end distance of neutral
polymers. For our issue, Yethiraj shows that in the limit
$L/\ell_e\to\infty$ the Gibbs-Bogoliubov approach and the uniform
expansion method applied to the end-to-end distance lead to the same
self-consistent equation~\cite{Yethiraj}. Clearly, both variational approaches
lead to similar results for a given observable. The important point is to
carefully consider, in both approaches, the behaviour at small length scales.

This method is applied both in the limit $\ell_0\ell_B/A^2\ll
1$ and at the so-called non-asymptotic limit, $\ell_b\ell_0/A^2\sim1$, when the
blob size is of the order of the bare persistent length
$\xi\simeq\ell_0$~\cite{Thirumalai1,Thirumalai2}. In the first case, they find
the KK result~\cite{Thirumalai2} but  in
their calculation they approximate the chain by a rod on a scale of the
order of $\ell_e$ (thus factorizing the statistical weights, see Eq.~(2.18) in
Ref.~\cite{Thirumalai2}). Moreover they skip \textit{a priori} the integrals on
long length scales, $s>\ell_e$, in the calculation, arguing that they lead to an
effective excluded-volume interaction. Hence, the dependence of the persistence
length on $N$ is eliminated. In the non-asymptotic limit, Ha and
Thirumalai use a simple Gaussian variational weight used for neutral and
flexible polymers and find
$\ell_e\simeq(\ell_0\ell_B/A^2)^{1/2}\kappa^{-1}$. However, as we explained in
the preceding section, this choice of Hamiltonian leads to a divergent entropic
term. To circumvent this problem, 
the evaluation of the statistical averages is done by imposing a
lower cutoff at $q\simeq\ell_e^{-1}$. We believe that this approximation leads
to the scaling $\ell_e\sim\kappa^{-1}$.

As shown by Muthukumar \textit{et al.}~\cite{Muthu2,Ghosh} the uniform expansion
method applied to $\langle R^2\rangle$ cannot be used to infer the scaling of
the persistent length. Indeed, to make the variational equation tractable (see
Eq.~(3.38) of Ref.~\cite{Muthu2}), they approximate the variational kernel
$h(\omega)$ (noted $l_1(q)$ in their paper) by the effective persistence length
$\ell_e$ [valid in the limit $\omega\to 0$, as it can be seen from
Eq.~(\ref{hw})] and they evaluate the integrand for $\omega\simeq2\pi/L$.
Hence, these approximations are valid in the limit of large length scales and
this method is powerful to get the large scale behaviour~: they indeed
recover the swelling exponent in the infinite limit $N\to\infty$, $\langle
R^2\rangle\sim N^{6/5}$ due to effective excluded volume interactions (which are
the screened electrostatic interactions if $\kappa^{-1}$ is non zero). But the
extrapolated persistence length from $\langle R^2\rangle\sim \ell_p N$ is
$\ell_p\sim N^{1/5}$, and does not correspond to the mechanical persistence
length. In this approach, the small scale behaviour is, in a sense, averaged
out, and all the electrostatics contribute to excluded volume interactions.

The uniform expansion can be used to calculate next-leading corrections in an
asymptotic expansion of the persistence length $\ell_e$ of the order of
$\mathcal{O}((\kappa\xi)^{-1}$. But in this case, it must be applied to to the
observable conjugated to the variational parameter, $\ell_e$, namely the bending
energy $\int \mathrm{d}s\,\mathbf{\ddot{r}}^2(s)$, and not to the end-to-end
distance. This could be valuable near the crossover, $\kappa\xi\sim 1$, between
the Gaussian-persistent regime and the Gaussian regime.

In two recent papers~\cite{Everaers,Nguyen}, simulations on long
polyelectrolytes (up to $N=4096$) show a behaviour consistent with the KK
formula~(\ref{KK}). Discrimination between $\ell_e\sim\kappa^{-1}$ and
$\ell_e\sim\kappa^{-2}$ is clearly seen only for very long polymers which is
one of the reason why the available experimental data are not very conclusive on
this point. Furthermore, in experiments there are other effects such as
counterion condensation and electrostatic correlations which are not included on
the Debye-H\"uckel level.


\section{Concluding remarks}


Our calculation is done within three important assumptions~: i) we study an
infinitely long chain ($N\to\infty$), ii) we do not consider excluded volume
interactions at large length scales and, iii) we take into account electrostatic
interactions at the linear level [a Debye-H\"uckel potential is used in
equation~(\ref{H})]. Within these assumptions, we show that the electrostatic
contribution to the persistence length, $\ell_e$, is proportional to
$\kappa^{-2}$, both for flexible (KK result), Eq.~(\ref{KK}), and stiff polymers
(OSF result), Eq.~(\ref{OSF}). This result is very robust (we tried a different
variational choice shown in the Appendix which leads to the same formula) and is
in contradiction with several theoretical papers where a $\kappa^{-1}$
dependence was found for flexible polymers. We show that this difference can be
explained by the use of a cutoff at small length scales in those works.

Obviously, a full comparison with experimental measurements on very long
chains would be necessary to validate this variational approach. Comparison
with experimentally determined persistence lengths of polyelectrolytes has been
done in several articles~\cite{Schmidt,Netz,Everaers}. Essentially, some
experimental papers raised some doubts about the validity of the OSF result and
proposed a law in $\ell_e\sim\kappa^{-1}$~\cite{Reed}. This law has been found
using approximate formulas for the radius of gyration to deduce the persistence
length, and the discrepancy can be due to an incomplete theory used in the data
analysis, as shown by Ghosh \textit{et al.}~\cite{Ghosh}.

More importantly, a deviation from the Debye-H\"uckel-type theory used in our
paper arises due to non-linear electrostatic effects. For highly charged
polymers (especially in the persistent regime), the renormalization of the bare
charge of the chain associated with the condensation of counterions (known as
the Manning condensation~\cite{Manning}) also influences the dependence of
$\ell_e$ on the Debye screening length. A recent study of this effect using a
variational approach applied to a charged cylinder~\cite{Netz2} shows that the
effective charge density along the polymer follows $A\sim \kappa^{-0.3}$ which
would lead to a modified power law $\ell_e\sim\kappa^{-1.4}$. We are currently
studying this effect for a semi-flexible polyelectrolyte.

Finally, we hope that this simple variational approach can be extended to the
issue of the electrostatic persistent length of polyelectrolytes in semi-dilute
solutions by combining this type of variational approach with the random-phase
approximation to account for monomer-monomer correlations and screening induced
by neighboring polymers~\cite{Muthu2,Ghosh,Yethiraj,Donley1,Donley2}.


\section*{Appendix}


In this appendix, we show that the KK result is also found with a piece-wise
continuous tangent-tangent correlation function which exhibits the same
behaviour as the modified WLC model at the asymptotic limits $|s|\ll
n^{1/2}\ell_e$ and $|s|\gg n^{1/2}\ell_e$. The advantage of this formulation is
that the interaction term, Eq.~(\ref{Hint}) is easily computed, and therefore we
can show that the large-scale term in the integrals is negligible in the
determination of the persistence length. Moreover, we thus prove that our result
for the salt-dependence of the persistence length is robust and insensitive to
the detailed choice of the variational correlation function.

Our variational choice for the squared monomer separation is
\begin{equation}
G_{\mathrm{pw}}(s) = \left\{
\begin{array}{lcr}
\ell_0 |s| +\frac{s^2}{n}\left(1-\frac13\frac{|s|}{n^{1/2}\ell_e}\right)
& \mathrm{for} & |s|<n^{1/2}\ell_e \\
\left(\ell_0+\frac{\ell_e}{n^{1/2}}\right)|s| - \frac{\ell_e^2}{3}
& \mathrm{for} & |s|>n^{1/2}\ell_e
\end{array}
\right.
\label{G}
\end{equation}
where ``pw" refers to piece-wise. For $|s|<n^{1/2}\ell_e$, the first line of
Eq.~(\ref{G}) is a sum of a linear term (Gaussian behaviour at small length
scales) and a term which imposes almost a rod-like behaviour for arc lengths
ranging in $n\ell_0<|s|<n^{1/2}\ell_e$. These linear and quadratic terms in the
first line of Eq.~(\ref{G}) are of equal magnitude for $s=n\ell_0$, the number
of monomers in a Gaussian blob at the onset of the persistent behaviour. For
$|s|>n^{1/2}\ell_e$, the statistics is again Gaussian, as expected for a
semi-flexible chain at large length scales.

The tangent-tangent correlation function is thus computed through
$\langle\mathbf{\dot{r}}(s)\mathbf{\dot{r}}(0)\rangle_{\mathrm{pw}} =
\frac12G_{\mathrm{pw}}''(s)$
\begin{equation}
\langle\mathbf{\dot{r}}(s)\mathbf{\dot{r}}(0)\rangle_{\mathrm{pw}}= \left\{
\begin{array}{lcr} \ell_0\delta(s) +
\frac{1}{n}\left(1-\frac{|s|}{n^{1/2}\ell_e}\right) & \mathrm{for} &
|s|<n^{1/2}\ell_e \\ 0 & \mathrm{for} & |s|>n^{1/2}\ell_e
\end{array}
\right.
\label{choice}
\end{equation}

\begin{figure}[htb]
\begin{center}
\includegraphics[height=6cm]{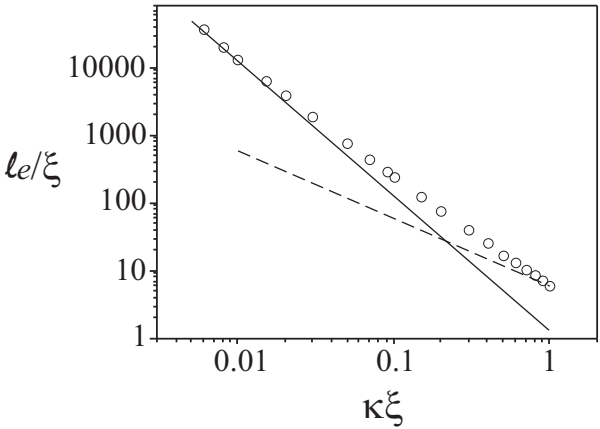}
\end{center}
\caption{Same plot as in Figure~\ref{minim} with a piece-wise variational
squared monomer separation function $G_{\mathrm{pw}}$}. \label{minim2}
\end{figure}

The function $h$, defined in Eq.~(\ref{hdef}), is then
\begin{equation}
h_{\mathrm{pw}}(\omega) = \ell_0\left[1+\frac{2}{\ell_e\xi}
\frac{1-\cos(\omega n^{1/2}\ell_e)}{n\omega^2}\right]
\end{equation}
We find exactly the same elastic part, Eq.~(\ref{Hel}), as in the modified WLC
model and for the entropy term we have
\begin{equation}
\beta\left(F_{\mathrm{pw}}-F_{\mathrm{uv}}\right)=
-\frac{\xi}{4\pi\ell_e}\frac{N}{n} \int_0^{\infty}\mathrm{d}x
\ln\left(1+2\frac{\ell_e}{\xi}\frac{1-\cos x}{x^2}\right) \label{entropypw}
\end{equation}
where $F_{\mathrm{uv}}$ is given in Eq.~(\ref{Fuv}). The interaction term is
found to be
\begin{eqnarray}
\frac{n}{N}\beta\langle\mathcal{H}_{\mathrm{int}}\rangle_{\mathrm{pw}}
&=&\int_0^1 \mathrm{d}x \left\{ \sqrt{\frac{6}{\pi}}
\frac{1}{\sqrt{x\xi/\ell_e + x^2 - x^3/3}}\right.\label{Hintpw}\\
&-&\left.\kappa\ell_e \exp\left[\frac{(\kappa\ell_e)^2}{6}
\left(\frac{x\xi}{\ell_e}+x^2 -\frac{x^3}{3}\right)\right]\right. \nonumber\\
&\times& \left.\mathrm{erfc}\left[\frac{\kappa\ell_e}{\sqrt{6}}\left(
\frac{x\xi}{\ell_e}+x^2-\frac{x^3}{3}\right)^{1/2}\right] \right\} \nonumber\\
&+& \frac{6}{\kappa(\xi+\ell_e)} \exp\left[\kappa^2\ell_e\xi\left(\frac16+\frac
{\ell_e}{9\xi}\right)\right] \nonumber\\
&\times& \mathrm{erfc}\left[\kappa\sqrt{\ell_e\xi} \left(\frac16+\frac
{\ell_e}{9\xi}\right)^{1/2} \right]  \nonumber
\end{eqnarray}
The numerical minimization of Eq.~(\ref{Fvar}) is shown in Fig.~\ref{minim2} and
shows that the KK result is also found with this variational approach.

\begin{figure}[htb]
\begin{center}
\includegraphics[height=4cm]{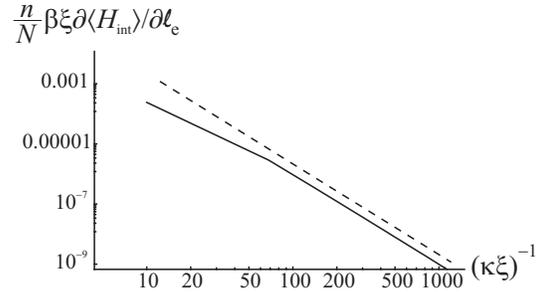}
\end{center}
\caption{First derivative of the interaction term per electrostatic
blob,$\frac{n}{N}\beta\xi\partial\langle \mathcal{H}_{\mathrm{int}}
\rangle_0/\partial\ell_e\vert_{\ell_e=(\kappa^2\xi)^{-1}}$, vs. $\kappa\xi$ when
$\ell_e$ is set to follow the KK law, $\ell_e\simeq(\kappa^2\xi)^{-1}$ in a log-log scale.
The dotted line is the power law $(\kappa\xi)^3$.} \label{Hint2fig}
\end{figure}

We have checked numerically that the entropy per electrostatic blob given by
Eq.~(\ref{entropypw}) has the same dependence in $-\sqrt{\xi/\ell_e}$ for
$\ell_e/\xi\to\infty$ as for the modified WLC model [Eq.~(\ref{entropymwlc})].
The last term of equation~(\ref{Hintpw}) corresponding to the large scale
behaviour, i.e. for $|s|>n^{1/2}\ell_e$, can easily be evaluated using the
saddle point expansion and varies like $\sim (\kappa\ell_e)^{-2}$. By applying
exactly the same routine as in Section~3, we show (Fig.~\ref{Hint2fig})
\begin{equation}
\frac{n}{N}\beta\langle \mathcal{H}_{\mathrm{int}}\rangle_0
+\ln(\kappa\xi) \simeq \frac{1}{\kappa\ell_e}
\end{equation}
which yields the same result as in Section~3. We thus note that the
electrostatic energy at small length scales, $|s|<n^{1/2}\ell_e$, is larger in
the limit $\ell_e/\xi\to\infty$ and governs the dependence of $\ell_e$ with
$\kappa$.


\acknowledgement{This work was financially supported by the Alexander von
Humboldt Foundation.}


\end{document}